\def\year{2020}\relax
\documentclass[letterpaper]{article} 
\usepackage{aaai20}  
\usepackage{times}  
\usepackage{helvet} 
\usepackage{courier}  
\usepackage[hyphens]{url}  
\usepackage{graphicx} 
\usepackage{booktabs}
\usepackage{float}
\usepackage{listings}
\usepackage{amsfonts,amssymb}
\usepackage{amsmath}
\usepackage{mathtools}

\DeclareMathOperator*{\argmin}{arg\,min}
\newfloat{lstfloat}{htbp}{lop}
\floatname{lstfloat}{Listing}
\usepackage{graphicx}  
\newcommand{\citet}[1]{\citeauthor{#1} \shortcite{#1}} \newcommand{\citep}{\cite} 
\begin{document}
\title{Learning Feature Interactions with Lorentzian Factorization Machine}

\author{
Canran Xu\textsuperscript{\rm 1}\thanks{Equal contribution.}, Ming Wu\textsuperscript{\rm 2}\footnotemark[1]\thanks{Work done during internship at eBay Inc.}\\
\textsuperscript{\rm 1}eBay Inc.\\
\textsuperscript{\rm 2}Department of Electrical Engineering, Tongji University, China\\
canxu@ebay.com, 1730683@tongji.edu.cn
}
\maketitle
\begin{abstract}
\begin{quote}
Learning representations for feature interactions to model user behaviors is critical for recommendation system and click-trough rate (CTR) predictions. Recent advances in this area are empowered by deep learning methods which could learn sophisticated feature interactions and achieve the state-of-the-art result in an end-to-end manner. These approaches require large number of training parameters integrated with the low-level representations, and thus are memory and computational inefficient. In this paper, we propose a new model named ``LorentzFM'' that can learn feature interactions embedded in a hyperbolic space in which the violation of triangle inequality for Lorentz distances is available. To this end, the learned representation is benefited by the peculiar geometric properties of hyperbolic triangles, and result in a significant reduction in the number of parameters (20\% to 80\%) because all the top deep learning layers are not required. With such a lightweight architecture, LorentzFM achieves comparable and even materially better results than the deep learning methods such as DeepFM, xDeepFM and Deep \& Cross in both recommendation and CTR prediction tasks.
\end{quote}
\end{abstract}

\lstset{ %
  language=python,                            
  basicstyle=\footnotesize\ttfamily,     
}

\section{Introduction}\label{section: introduction}
Recommendation system and ads click-trough rate (CTR) prediction both have profound impact on web-scale big data, and have been received a lot of attention in both academia and industrial community. In these systems, the critical problem is to predict the probability of clicks, which will be used to rank the candidate items or ads. 

These problems have been successfully tackled by supervised learning methods, in which user profiles and item attributes are used as input features. The challenge is that, the data for web-scale recommendation systems is mostly discrete and categorical, resulting in an extreme large and sparse feature space to optimize. Feature sparsity of this kind is commonly handled by constructing feature interactions which are modeled by inner product of their low dimensional representations. In this so-called factorization machine model \cite{FM}, the naive inner product is not expressive enough for spurious or implicit feature interactions. Therefore, higher-order factorization machine \cite{HOFM} is proposed to learn higher-order feature interactions efficiently. On the other hand, deep learning based approaches have successfully enhanced the performance significantly because sophisticated feature interactions can be captured. Besides, another merit of the deep learning methods is that they can reduce human efforts in feature engineering. To name a few, DeepFM \cite{DeepFM} integrates the deep neural networks on top of the factorization machine, and the Deep \& Cross neural network (DCN) \cite{DCN} takes outer product of features at the element wise and vector wise level such that higher order feature interactions can be learned in an automatic and explicit fashion. In DCN, the order of interactions is determined by the number of multi-layer perceptron (MLP) layers. However, by using multiple layers of MLPs it is inefficient to learn multiplicative feature interactions. Therefore, in xDeepFM \cite{Xdeepfm} it is proposed to use convolution neural network to replace MLP.

In this paper, we re-examine the building block for these models: the inner product between feature representations. To this end, we propose to learn the low-dimensional representations in a hyperbolic geometry equipped with Lorentz distance, such that the triangle inequality for feature vectors could be violated. Our work is inspired by the work on collaborative metric learning (CML) \cite{cml}, where the inner product is replaced by the distance between feature vectors, but following a completely different path. As opposed to the work of CML, in which the authors argue that triangle inequality in a Euclidean space should be strictly obeyed, we propose to take advantage of the sign of the triangle inequality. Specifically, we construct our score function for feature interactions not by the inner product or distance between feature vectors, but by examining the triangle formed by them in a hyperbolic space. This reason for our approach is two-folded: (1) a hyperbolic space is intrinsically a lot more capacious than the Euclidean space; (2) the proposed score function will provide a robust objective function to learn fine-grained feature interactions. As a result, the embedding learned by this score function are therefore ready to use directly, just as the vanilla factorization machine. Our experiments on several benchmark datasets show that, even without using any deep learning layers on top of the embedding layer, the proposed approach will achieve the state-of-the-art performance with reducing up to 82.9\% in training parameters and 69.7\% in training time compared to the strong deep learning baselines such as DeepFM, DCN and xDeepFM.

\subsection{Our Contributions}
Overall, the primary contributions of this paper can be summarized as follows:
\begin{itemize}
    \item We propose to learn feature interactions with a score function measuring the validity of triangle inequalities. This score function is bounded and dimension agnostic, which opens up a novel viewpoint on learning the fine-grained structure of interacting features.
    \item We conduct extensive experiments on a wide range of real-world datasets. Our results demonstrate that LorentzFM achieves state-of-the-art performance, compared to existing deep learning approaches.
    \item We thoroughly investigate the number of training parameters and training time to understand its cost in resource.
\end{itemize}

\section{Why Learning Triangle Inequality?}\label{sec:background}
In this section, we first provide some technical background about hyperbolic geometry and Lorentz distance. With these definitions,  we will show that Lorentz distances for a triangle could violate triangle inequality. The motivation is to show how it is related to the metric learning in Euclidean space.

\subsection{Hyperbolic Geometry and Lorentz Distance}
Hyperbolic geometry aims to study non-Euclidean space with a constant negative curvature. Due to its negative curvature, hyperbolic geometry has very different properties compared to the Euclidean geometry. 

First, the circumference and area of a circle in the hyperbolic space grow exponentially with the radius, as opposed to the linear and quadratic growth rate in Euclidean space. Therefore, the capacity of embeddings in the hyperbolic space of the same upper bound in radius is much larger than its Euclidean counterpart. Secondly, the triangle inequality Eq. \eqref{eq:triangle_ineq} with Lorentz distance defined in Eq. \eqref{eq:lorentz_distance} defined therein can be violated. This property enables us to characterize the pairwise relation between points in the hyperbolic space by the sign of the inequality.

There are several important models of hyperbolic space for computation purpose: the Poincar\'{e} ball model, the hyperboloid model, the Klein model and so on. These models all describe the same geometry and can be connected by transformations that preserve geometric properties of the hyperbolic space such as isometry. In the following, we only introduce the technical basics of the hyperboloid model, which will be used throughout the paper.

\subsubsection{Hyperboloid Model} 
Let us define the Lorentzian inner product between $\mathbf{u}$, $\mathbf{v}\in\mathbb{R}^{n+1}$ as
\begin{equation}\label{eq:lorentz_inner_prod}
    \langle \mathbf{u}, \mathbf{v} \rangle_{\mathcal{L}} 
    = -u_0 v_0 + \sum_{i=1}^nu_n v_n.
\end{equation}
The hyperboloid of dimension {\em n},  $\mathcal{H}^{n,\beta}\subseteq\mathbb{R}^{n+1}$ consists of the following set of points:
\begin{equation}\label{eq:hyperboloid}
    \mathcal{H}^{n,\beta} = \left\{\mathbf{x}\in\mathbb{R}^{n+1} : ||x||_{\mathcal{L}}^2 = -\beta, x_0 \geq \beta\right\},
\end{equation}
where $||\mathbf{x}||_{\mathcal{L}}^2 = \langle \mathbf{x}, \mathbf{x}\rangle_{\mathcal{L}}$ is the Lorentzian norm of the vector $\mathbf{x}$.  With this definition, for every vector $\mathbf{x}\in\mathcal{H}^{n,\beta}$, the component $x_0$ is not a free parameter, and is given by
\begin{equation}\label{eq: x_0}
    x_0 = \sqrt{\beta + \sum_{i=1}^n x_i^2}.
\end{equation}
The associated geodesic distance between two points is given as
\begin{equation}\label{eq:geodesic}
    d_{\ell}(\mathbf{u}, \mathbf{v}) = {\rm arccosh}(-\langle \mathbf{u}, \mathbf{v} \rangle_{\mathcal{L}} ).
\end{equation}
Note that the origin vector of the hyperboloid model $\mathcal{H}^{n,\beta}$ is $\mathbf{0}=(\beta,0,\dots,0)$, and its Lorentzian inner product with vector $\mathbf{x}$ is simply $\langle \mathbf{0}, \mathbf{x}\rangle_\mathcal{L} = -x_0 \leq \beta$. 

When $\beta=1$, the model is called a \textit{unit} hyerboloid model, which will be used throughout the paper. Without introducing any confusions, we will simply call it hyperboloid model and use $\mathcal{H}^{n}$ to denote $\mathcal{H}^{n,1}$.

\subsubsection{Lorentz Distance} 
\begin{figure}[t]
\begin{center}
\includegraphics[width=1\columnwidth]{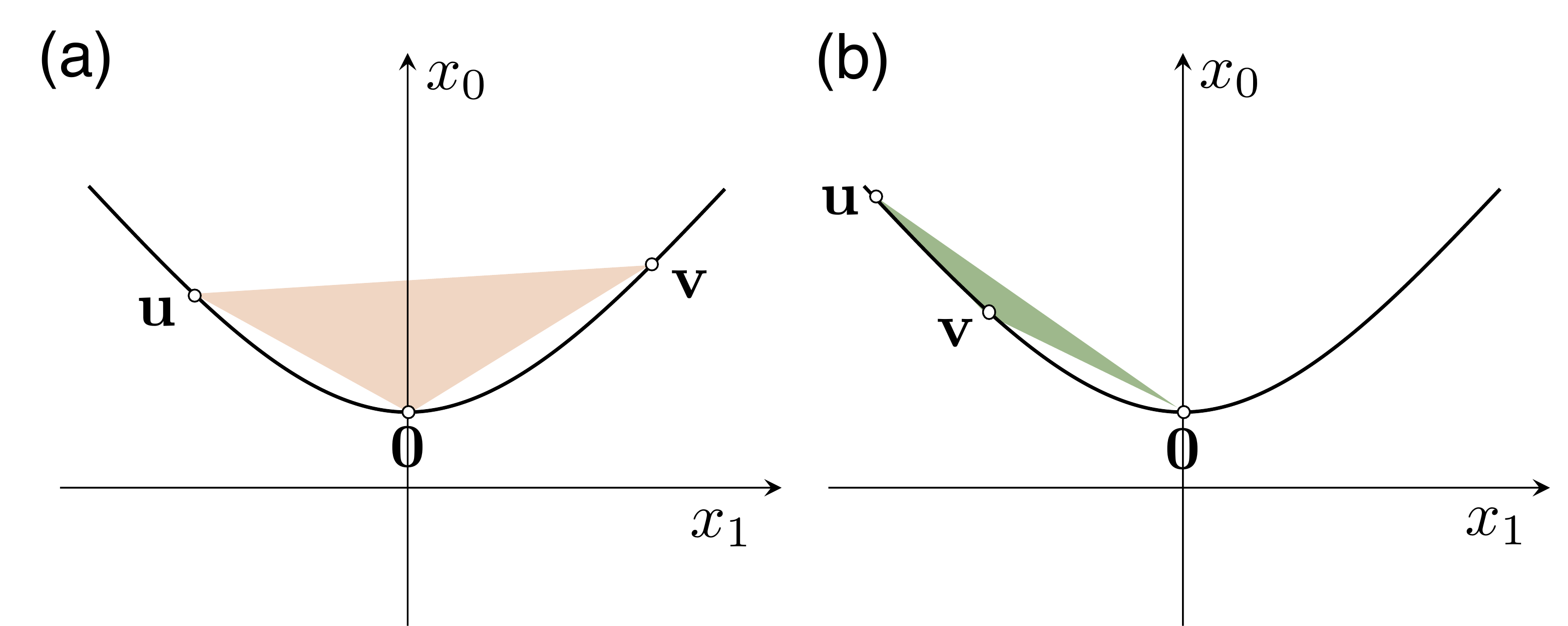}
\end{center}
\caption{Visualization of the hyperboloid model $\mathcal{H}^2$. In this model, triangles could (a) violates Lorentz triangle inequality and (b) obeys triangle inequality for points on the hyperboloid.} \label{figure:triangle}
\end{figure}

The squared Lorentz distance, or Lorentz distance in short, between $\mathbf{u}$, $\mathbf{v}\in\mathcal{H}^n$ is given by
\begin{equation}\label{eq:lorentz_distance}
     d^2_\mathcal{L}(\mathbf{u}, \mathbf{v}) = ||\mathbf{u} - \mathbf{v}||_{\mathcal{L}}^2 = -2 - 2\langle\mathbf{u}, \mathbf{v}\rangle_{\mathcal{L}}.
\end{equation}
It satisfies almost all axioms of Euclidean geometry but the triangle inequality, which is one of the most crucial geometric property with positive definitive Riemannian metric. It states that for any three points, $\mathbf{x}, \mathbf{y}$ and $\mathbf{z}$, any two pairwise distances $d(\cdot, \cdot)$ should be greater or equal to the remaining pairwise distance:
\begin{equation}\label{eq:triangle_ineq}
    d(\mathbf{x}, \mathbf{y}) \leq d(\mathbf{x}, \mathbf{z}) + d(\mathbf{z}, \mathbf{y}).
\end{equation}
In the hyperbolic space, the geodesic distance defined in Eq. \eqref{eq:geodesic} between three points also satisfies this property. However, the inequality could be violated with Lorentz distances, because the Riemannian metric is negative.  Consider the triangle formed by the origin $\mathbf{0}$ and other two points $\textbf{u}$ and $\textbf{v}$ in Figure \ref{figure:triangle}. When the two points are far apart on the opposite side of $x_1$ axis, the triangle inequality is violated. On the other hand, the triangle inequality holds if the two points are on the same side of $x_1$ axis. 

\subsection{Learning Triangle Inequalities}

As pointed out in \citet{cml}, learning the distance in the embedding space rather than inner product has advantages to learn a fine-grained embedding space that could capture the representation not only for item-user interactions, but  also for item-item and user-user distances. Essentially, the so-called metric learning scheme is blessed by the constraint of triangle inequality.

\begin{figure}[ht]
\begin{center}
\includegraphics[width=1\columnwidth]{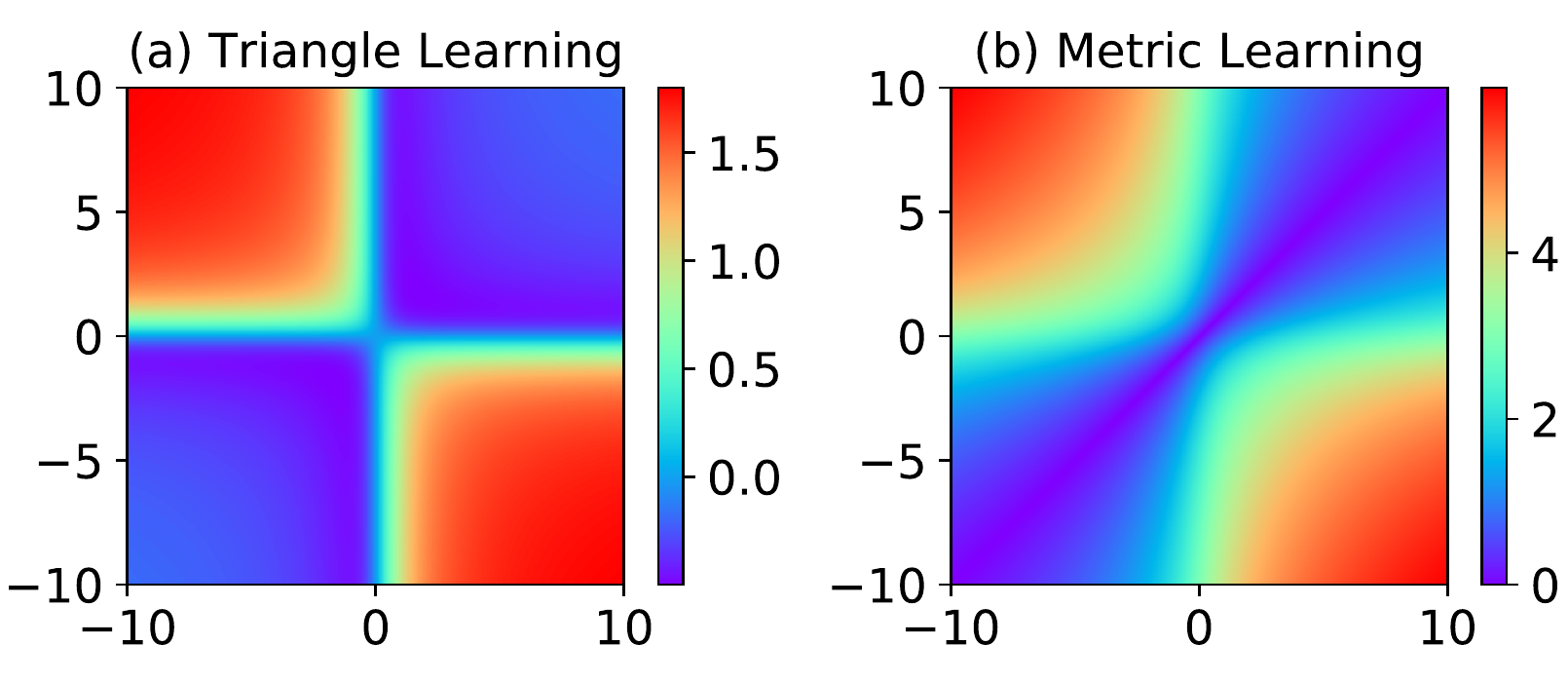}
\end{center}
\caption{The 2D landscape for the score function in the (a) triangle learning scheme ({\em e.g.}, Eq. \eqref{eq:triangle_inequality}) and (b) metric learning scheme using geodesic distance in a 2D hyperboloid model. } \label{figure:landscape}
\end{figure}

In contrast to collaborative metric learning scheme, we argue that the feature interaction between two points can be learned by the sign of the triangle inequality for Lorentz distance, instead of using the distance itself. Formally, our target score function is written as
\begin{equation}\label{eq:triangle_inequality}
    \mathcal{S}(\mathbf{x}, \mathbf{y}) = \frac{ d^2_{\mathcal{L}}(\mathbf{x}, \mathbf{y}) - d^2_{\mathcal{L}}(\mathbf{x}, \mathbf{0}) - d^2_{\mathcal{L}}(\mathbf{0}, \mathbf{y})}{\langle \mathbf{0}, \mathbf{x} \rangle_{\mathcal{L}} \langle \mathbf{0}, \mathbf{y}\rangle_{\mathcal{L}}},
\end{equation}
where $\mathbf{0} = (1, 0, \cdots, 0)$ is the origin. The nominator is simply the difference between the two sides of the triangle inequality, and the denominator is used to bound the score function. 

The advantage of using Eq. \eqref{eq:triangle_inequality} is that, the function is bounded in $[-0.5, 2]$ in all dimensions. Therefore, the score function is free from the curse of dimensionality, compared to the collaborative metric learning scheme. To illustrate the landscape of the function, we plot Eq. \eqref{eq:triangle_inequality} for two points in 2D  in Figure \ref{figure:landscape} (a). Because each point on the 2D hyperboloid has only one free parameter, we plot the landscape of the score function by setting the $x$-axis and $y$-axis corresponding to the free parameter of each point respectively in a 2D plane. We also plot the landscape for collaborative metric learning scheme using the geodesic distance defined in Eq. \eqref{eq:geodesic} in Figure \ref{figure:landscape} (b). By comparison, we can observe that the landscape for our approach is smooth and bounded, but the score function for collaborative metric learning scheme is unbounded. The bounded property is promising because the embedding vectors are free to be far away from the origin with a smooth growth in the score function.

\section{Lorentzian Factorization Machine}

In this section, we first give an overview of our proposed Lorentzian Factorization Machine or {\em LorentzFM} for short, named after learning the triangle inequality equipped with Lorentz distance. Next, we describe each component of the model in some greater details.
Finally, we present a detailed description of how to optimize the model.

\begin{figure}[ht]
\begin{center}
\includegraphics[width=0.99\columnwidth]{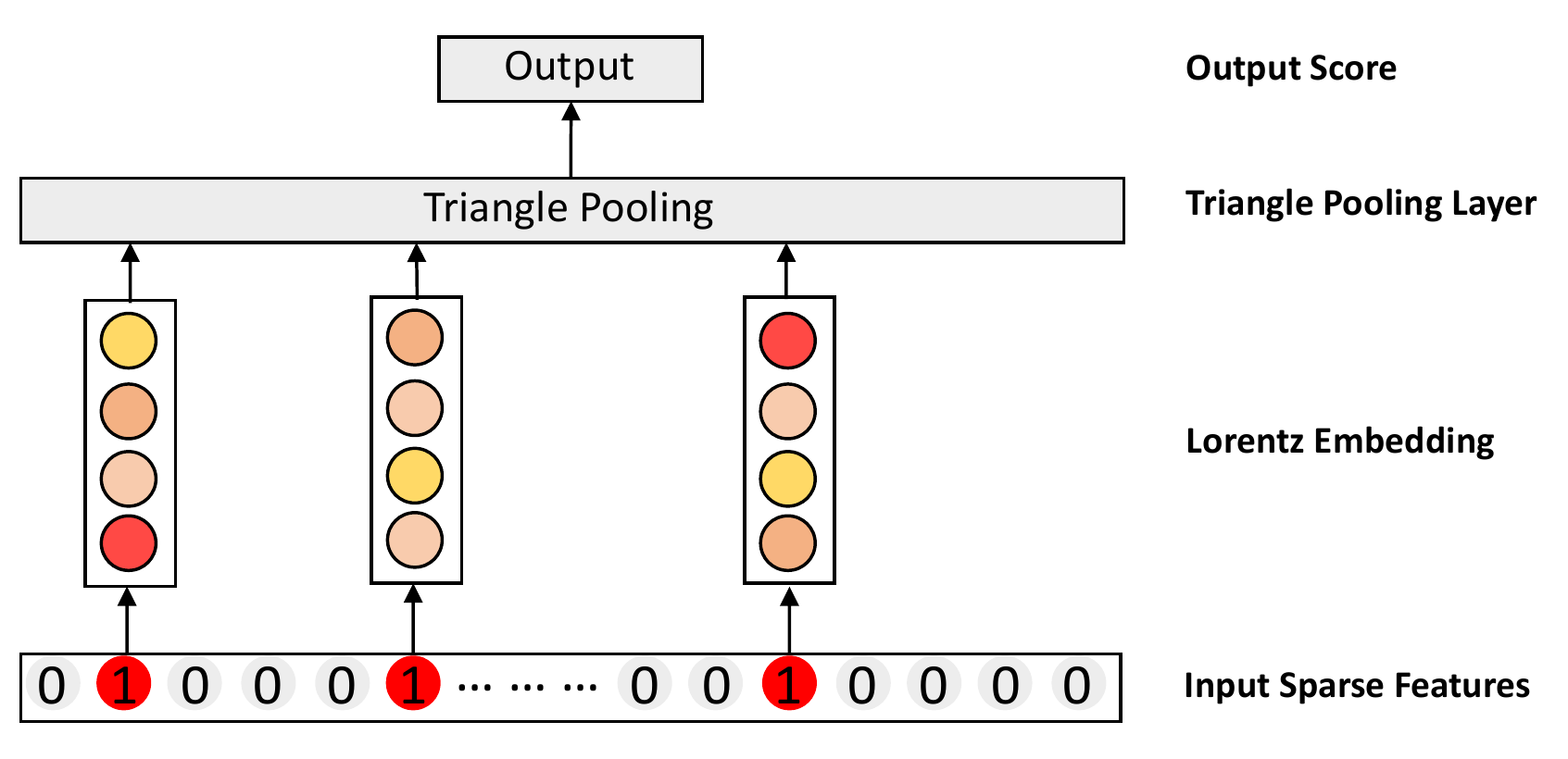}
\end{center}
\caption{The architecture of LorentzFM. It comprises an input layer, a Lorentz embedding layer and a triangle pooling layer to aggregate feature interactions.} \label{figure:model_architecture}
\end{figure}

\subsection{Overview}

The goal of our approach is to map the original sparse features into a low-dimensional space. As shown in Figure \ref{figure:model_architecture}, our purposed method takes the sparse feature vector $\mathcal{V}_x$ as input, followed by a Lorentz embedding layer that projects all features into the same hyperbolic space. Next, we feed the embeddings of all fields into a novel triangle pooling layer, which is coined as an aggregation function of all feature pairs to measure the soft ``validness'' of triangle inequality in overall. Unlike the recent state-of-the-arts neural architectures built upon Euclidean embeddings, LorentzFM does not need any extra parameters. In particular, for given sparse input $\mathcal{V}_x$, the output of the pooling layer is the model output score:
\begin{equation}\label{eq: LFM}
    \hat{S}_{LFM}(\mathcal{V}_x) = \sum_{i,j=1,i\neq j}^d \mathcal{T}(\textbf{v}_i, \textbf{v}_j)x_i x_j,
\end{equation}
where $\textbf{v}_i, \textbf{v}_k \in \mathcal{H}^{n}$ are embedding vector for each input feature field, and $\mathcal{T(\cdot, \cdot)}$ is the feature interaction function. Although formally in Eq. \eqref{eq: LFM} the linear term is missing, it actually reappears in the pooling function, $\mathcal{T(\cdot, \cdot)}$, as shown later.

\subsection{Lorentz Embedding Layer}
The embedding layer is a lookup table to project sparse features to low-dimensional dense vectors in the Lorentz space. Formally, let $\textbf{v}^k\in\mathcal{H}^{n}$ be the embedding vector for the $k$-th feature, whereas the 0-th component is given by the constraint as Eq. \eqref{eq: x_0}.

In some cases, categorical features can be multi-valued. For example, the genre of movie ``Titanic'' could be either ``Drama'' or ``Romance''. We use multiple fields for these categorical features under a universal encoding and pad them with ``unknown'' tags to ensure each sample is aligned in feature dimension.

\subsection{Triangle Pooling Layer}
We then feed the embedding vectors into the triangle pooling layer, which is an aggregation function that converts a set of embedding vectors to one vector:
\begin{equation}\label{eq:triangle_pooling}
\begin{aligned}
    \mathcal{T}(\textbf{u}, \textbf{v}) &= \frac{d^2_{\mathcal{L}}(\textbf{u}, \textbf{v}) - d^2_{\mathcal{L}}(\textbf{0}, \textbf{u}) - d^2_{\mathcal{L}}(\textbf{0}, \textbf{v})}{2 \langle\textbf{0}, \textbf{u} \rangle_{\mathcal{L}} \langle\textbf{0}, \textbf{v} \rangle_{\mathcal{L}}} \\
    &= \frac{1 - \langle\textbf{u}, \textbf{v} \rangle_{\mathcal{L}}  - u_0 - v_0}{u_0 v_0} \\
    &= \underbrace{\frac{1 - \langle\textbf{u}, \textbf{v} \rangle_{\mathcal{L}} }{u_0 v_0}}_\text{\clap{interaction term}} - \underbrace{\left( \frac{1}{u_0} + \frac{1}{v_0} \right)}_\text{\clap{linear term}}.
\end{aligned}
\end{equation}
In the second line we use the definition of Lorentz distance and Lorentz inner product. Owing to the normalization denominator, the linear term emerges, as shown in the last line.

\subsection{Objective and Learning}
The objective function to optimize both recommendation system and CTR prediction is the binary cross-entropy (BCE):
\begin{equation}
    \argmin_{\theta} \sum_{i} -y_i \log (p_i) - (1 - y_i) \log (1 - p_i),
\end{equation}
where $p_i$s are the probabilities for the $i$-th sample input vector $\mathcal{V}_{x}^{(i)}$ and computed as $p_i = \sigma(\hat{S}_{LFM}(\mathcal{V}_{x}^{(i)}))$, and $y_i$s are the true labels. Even though Bayesian Personalized Ranking (BPR) loss \cite{bpr} proves to be useful in common recommendation systems, we do not use it because the BCE loss is sign-sensitive, which is the desired property, while the BPR loss is not.

The parameters of our model are learned by using Riemanian stochastic gradient descent (RSGD) \cite{Rsgd}. As shown by \citet{lorentz_model}, the parameters are updated by the following form
\begin{equation}
    \theta_{t+1} = \exp_{\theta_t}(-\eta \,{\rm grad} \, f(\theta_t)),
\end{equation}
where ${\rm grad}\,f(\theta_t)$ is the gradient defined in the Riemannian manifold and $\eta$ is the learning rate. The Riemanian gradient is obtained by multiplying the gradient in the Euclidean space by the Lorentz metric, and then performing an orthogonal projection onto the tangent space spanned by the current parameter set. Finally, the parameter update is given by the following exponential map 
\begin{equation}
    \exp_{\theta_t} (\mathbf{x})=\cosh(||{\mathbf{v}}||_{\mathcal{L}}) \mathbf{x} + \sinh({||{\mathbf{v}}||_{\mathcal{L}}}) \frac{\mathbf{v}}{||
    \mathbf{v}||_{\mathcal{L}}}
\end{equation}
which maps a tangent vector $\mathbf{v}$ in the tangent space onto the Lorentz manifold. Details can be found in \citet{lorentz_model}.

Since the score function Eq. \eqref{eq:triangle_inequality} is bounded and dimension independent, it is not necessary to apply an $L_2$ regularization term over the embedding vector, because it will create a resisting gradient towards the origin of the hyperboloid model. 

\section{Experiment}\label{section:experiment}
In this section, we evaluate LorentzFM with four real-word datasets. We aim to answer the following research questions:
\begin{itemize}
    \item \textbf{RQ1:} Compared with the state-of-the-art deep learning based methods, how does our method perform?
    \item \textbf{RQ2:} What is the resource cost for LorentzFM compared to existing deep learning based methods?
    \item \textbf{RQ3:} How can we interpret feature interactions learned by LorentzFM?
\end{itemize}

\begin{table}[ht]
\begin{center}
\begin{tabular}{lcccc}
\toprule
 & {Steam} & {MovieLens} & {KKBox} & {Avazu$^{\dagger}$}\tabularnewline
 \midrule
\#Samples & 1.23M & 573K &  2.67M & 40.4M\tabularnewline
\midrule
\#User Fields & 1 & 5& 7& --\tabularnewline
\#Item Fields & 11 & 7& 6& --\tabularnewline
\#Fields & 12 & 12 & 12 & 21 \tabularnewline
\midrule
\#Users & 26.6K & 6.0K & 21.1K & --\tabularnewline
\#Items & 5.9K & 3.1K & 9.3K & --\tabularnewline
\#Features & 46.2K & 9.4K & 38.2K & 972K \tabularnewline
\midrule
Sparsity & 99.22\% & 96.85\% & 98.64\% & -- \tabularnewline
\bottomrule
\end{tabular}
\caption{Statistics of datasets after data preprocessing. $^{\dagger}$ In Avazu dataset there are no explicit indicators for item ID and user IDs, so we leave the corresponding rows blank.}\label{table: data_stats}
\end{center}
\end{table}

\begin{table*}[ht]
\centering
\begin{tabular}{lccccccccccc}
\toprule
 & \multicolumn{3}{c}{Steam} & \multicolumn{3}{c}{MovieLens} & \multicolumn{3}{c}{KKBox} & \multicolumn{2}{c}{Avazu}\tabularnewline
\cmidrule(lr){2-4}
\cmidrule(lr){5-7}
\cmidrule(lr){8-10}
\cmidrule(lr){11-12}
 & MRR & HR@10 & NDCG & MRR & HR@10 & NDCG & MRR & HR@10 & NDCG & AUC & Logloss\tabularnewline
 \midrule
FM & 0.1042 & 0.1898 & 0.2376 & 0.0638 & 0.1258 & 0.1966 & 0.0578 & 0.0998 & 0.1735 & 0.7582 & 0.3920  \tabularnewline
NFM & 0.1213 & 0.2286 & 0.2591 & 0.0723 & 0.1446 & 0.2082 & 0.0586 & 0.1086 & 0.1779 & 0.7641 & 0.3884 \tabularnewline
DeepFM & 0.1284 & 0.2408 & 0.2664 & 0.0780 & 0.1548 & 0.2143 & 0.0637 & 0.1140 & 0.1843 & 0.7674 & 0.3864 \tabularnewline
xDeepFM & 0.1228 & 0.2316 & 0.2615 & \textbf{0.0791} & \textbf{0.1570} & \textbf{0.2156} & 0.0639 & 0.1148 & 0.1846 & 0.7677 & 0.3865 \tabularnewline
DCN & 0.1213 & 0.2261 & 0.2598 & 0.0749 & 0.1512 & 0.2119 & 0.0644 & 0.1136 & 0.1848 & 0.7677 & 0.3863 \tabularnewline
\midrule
LorentzFM & \textbf{0.1362} & \textbf{0.2592} & \textbf{0.2746} & 0.0780 & 0.1533 & 0.2129 & \textbf{0.0733} & \textbf{0.1295} & \textbf{0.1938} & \textbf{0.7775} & \textbf{0.3828} \tabularnewline
\bottomrule
\end{tabular}
\caption{Best performance of each model with fixed embedding size 10 on the test set.} \label{table:performance}
\end{table*}

\subsection{Datasets and Preprocessing}
We use Steam, MovieLens and KKBox datasets to evaluate the recommendation task and Avazu dataset for the CTR task. The datasets are briefly described as follows.
\begin{itemize}
\item \textbf{Steam}\footnote{\url{http://cseweb.ucsd.edu/~jmcauley/datasets.html#steam_data}} is a dataset crawled from Steam database, which includes rich information such as users' playing hours, games' price, category and publisher \textit{etc}. We keep positive samples by setting the threshold of playing hours to 100.
\item \textbf{MovieLens}\footnote{\url{https://grouplens.org/datasets/movielens/1m/}} is a set of benchmark datasets for evaluating recommendation algorithms. We use the MovieLens 1M version and keep samples with ratings above 3. 
\item \textbf{KKBox}\footnote{\url{https://www.kaggle.com/c/kkbox-music-recommendation-challenge}} dataset is adopted from the WSDM cup 2018 Challenge provided by the music streaming service KKBox. This dataset includes demographic information about the users and structured meta data about the songs. The original dataset has a binary label and we only keep the positive samples. 
\item \textbf{Avazu}\footnote{\url{https://www.kaggle.com/c/avazu-ctr-prediction}} is an ads click-through dataset with more than 40 millions instances. It consists of 10 days of ad click-through data with only sparse features. 
\end{itemize}

For the three recommendation datasets we process the dataset as follows: (1) keeping only positive ratings as described above and filter the \{KKBox, Steam, MovieLens\} dataset to \{20, 20, 5\}-core setting; (2) randomly splitting \{10K, 10K\} samples for \{validation, testing\}, and the rest as training set. For the Avazu dataset, we remove the timestamp field and then replace features less than 5 times by a universal ``unknown'' tag.  We randomly select 80\% samples for training and, 10\% for validation and 10\% for testing. Data statistics are shown in Table \ref{table: data_stats}.

\begin{figure*}[ht]
\begin{center}
\includegraphics[width=1.8\columnwidth]{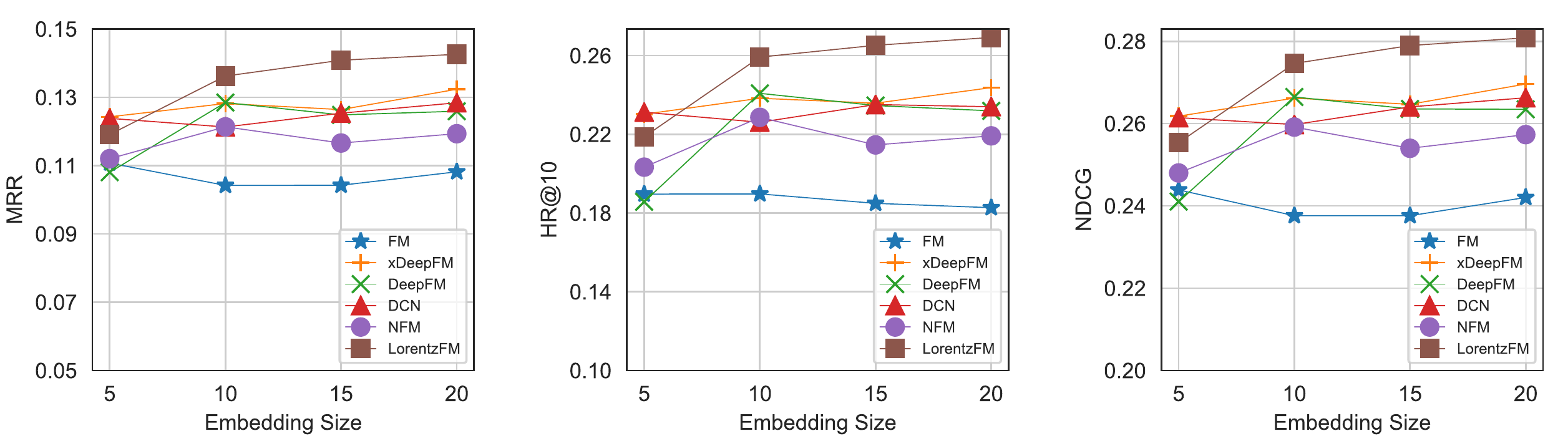}
\end{center}
\caption{Performance comparison on the test set {\em w.r.t.} different embedding sizes for the Steam dataset.} \label{figure:dim_performance}
\end{figure*}

\subsection{Experiment Settings}
\subsubsection{Evaluation Metrics}
For each positive user-item pair in the recommendation task, we calculate the following three ranking metrics to evaluate the performance over {\em all} unobserved samples:
\begin{itemize}
    \item \textbf{MRR} is the average of the reciprocal ranks of the test items by their scores over the entire ranking list.
    \item \textbf{HR@10} measures the relative orders among positive and negatives within the top 10 of the ranking list.
    \item \textbf{NDCG} accounts for the position of the test items by assigning higher scores to top ranks over the entire ranking list.
\end{itemize}
For the CTR prediction task, we use {AUC} (Area Under the ROC curve) and {logloss} (binary cross-entropy) for evaluation. AUC measures the probability that an instance prediction will be ranked a higher score to a randomly chosen negative item, and in contrast logloss measures the distance between the predicted score and the ground truth label for each sample.

\subsubsection{Compared Baselines}
We consider the following models are our baselines:
\begin{itemize}
\item \textbf{FM} \cite{FM} combines the second-order feature interaction as well as a first order term.
\item \textbf{NFM} \cite{NFM}  can learn implicit high-order features by stacking MLP on top of the FM model.
\item \textbf{DCN} \cite{DCN}  concatenates MLP and Cross Network, which takes outer product of concatenated feature vector at bit-wise level. It can extract both the implicit and explicit high-order features.
\item \textbf{DeepFM} \cite{DeepFM}  combines the FM model and deep MLP to get the low and high-order features.
\item \textbf{xDeepFM} \cite{Xdeepfm} takes out product at the vector-wise level and applies the convolution layers to get the high-order explicit and implicit features.
\end{itemize}

\begin{table*}[ht]
\begin{center}
\begin{tabular}{lcccccccc}
\toprule
 & \multicolumn{4}{c}{\# Training Parameters}
 & \multicolumn{4}{c}{Training Time per Epoch (seconds)} \tabularnewline 
 \cmidrule(lr){2-5}
\cmidrule(lr){6-9}
 & Steam & MovieLens & KKBox & Avazu & Steam & MovieLens & KKBox & Avazu \tabularnewline
 \midrule
FM & 507k & 103k & 420k & 10.68M & 124.0 & 82.4 & 648.3 & 312.1 \tabularnewline
NFM & 519k & 115k & 432k & 11.36M & 158.7 & 87.7 & 746.6 & 478.2 \tabularnewline
DeepFM & 717k & 313k & 634k & 11.10M & 159.2 & 87.1 & 693.3 & 450.7 \tabularnewline
xDeepFM & 865k & 498k & 808k & 11.46M & 391.0 & 223.4 & 1340.2 & 1459.8 \tabularnewline
DCN & 530k & 126k & 444k & 11.24M & 152.6 & 62.16 & 666.8 & 547.2 \tabularnewline
\midrule
LorentzFM & 415k & 85k & 344k & 8.75M & 160.7 & 67.6 & 664.2 & 522.2 \tabularnewline
$-\Delta$\%  & 21.8\%  & 82.9\% & 57.4\% & 22.1\% & -8.1\% & 69.7\% & 50.4\% & 4.6\% \tabularnewline
\bottomrule
\end{tabular}
\caption{Number of training parameters and training time per epoch according to the best performance in Table  \ref{table:performance}. The last row corresponds to the reduced percentage in training parameters or training time of LorentzFM, compared to the best performed baseline.}\label{table:cost_compare}
\end{center}
\end{table*}

\subsection{Training Details}
We implement LorentzFM and all baselines in PyTorch\footnote{\url{https://pytorch.org/}} \cite{pytorch}, and for each model the reported performance is given by performing grid search over the hyper-parameters on the validation set, with early stopping at 20 epochs.  The embedding size is set to 10 for all models for fair comparisons. 
For the recommendation task, for each-item pair in the training set, we randomly sample 10 unobserved items together with the given user as negative samples. MRR is the metric for early stopping. For the CTR task, we use the original binary labels for training, and logloss is monitored for early stopping .

\textbf{LorentzFM Details}  Burn-in period is set to 25 epochs. For the recommendation task, we tune the learning rate $\in$ [0.05, 0.1, 0.2, 0.3] with RSGD and batch size $\in$ [64, 128, 256, 512]. For the CTR prediction task, the learning rate is tuned $\in$ [0.1, 0.2, 0.3] and batch size is set to 4096. 

\textbf{Baselines Details}  The loss function is BPR loss for the recommendation task and BCE loss for the CTR task. We tune the dropout rate $p \in \{0,\,0.1,\dots,\,0.5\}$, the MLP layers $\in$ [[512, 256], [400, 400], [100, 100]], and the batch size $\in$ [64, 128, 256, 512]. 
Learning rate is set to 0.001 with Adam optimizer \cite{Adam}. We use an $L_2$ regularization with $\lambda=10^{-5}$. For the CTR prediction task, all hyper-parameters except for the embedding size are set identical to the original paper.

\subsection{Performance Comparison (RQ1)}
Table \ref{table:performance} reports our experiment results of each individual model on the four datasets with fixed embedding size 10. For each dataset, all the deep learning methods (NFM, DeepFM, xDeepFM and DCN) substantially outperform FM. It is expected because additional neural net layers built on top of the second-order feature interaction layer will incorporate higher-order feature interactions. The performance between these deep learning models are close upon fine-tuning over the hyper-parameters. Surprisingly, LorentzFM achieves even better results compared to these deep learning models significantly except for the MovieLens dataset. We observe that in MovieLens the number of sparse features is 5-100 times smaller than the other three datasets, which reflects that LorentzFM is especially powerful when the data is extremely sparse.

In order to understand the impact of embedding size on LorentzFM as well as the compared baselines, we conduct control experiments on the Steam dataset by holding the best hyper-parameter settings for each model in Table \ref{table:performance}, while varying the embedding size $\in[5,\,10,\,15,\,20]$. Figure \ref{figure:dim_performance} demonstrates that in general the performance of all models increases with embedding size except for FM. Note that the performance of LorentzFM surpasses over other deep learning methods at embedding size $\in [10,\,15,\,20]$.

\subsection{Resource Cost Comparison (RQ2)}
Table \ref{table:cost_compare} compares the amount of training parameters and the training time per epoch between LorentzFM and the best tuned baselines. In order to compare the training time fairly, we use the same batch size for each model. The number of training parameters for LorentzFM is 9 times feature size, because the zeroth component of each embedding vector is not a free parameter. In the last row of Table \ref{table:cost_compare}, we show the reduced percentage of the corresponding training parameters and training time when LorentzFM is compared with the best performing baseline. Because LorentzFM does not need any deep learning layers on top of the triangle pooling layer, it could reduce up to 82.9\% training parameters and 69.7\% in training time compared to xDeepFM.

\subsection{Case Study (RQ3)}

\begin{figure}[ht]
\begin{center}
\includegraphics[width=1.\columnwidth]{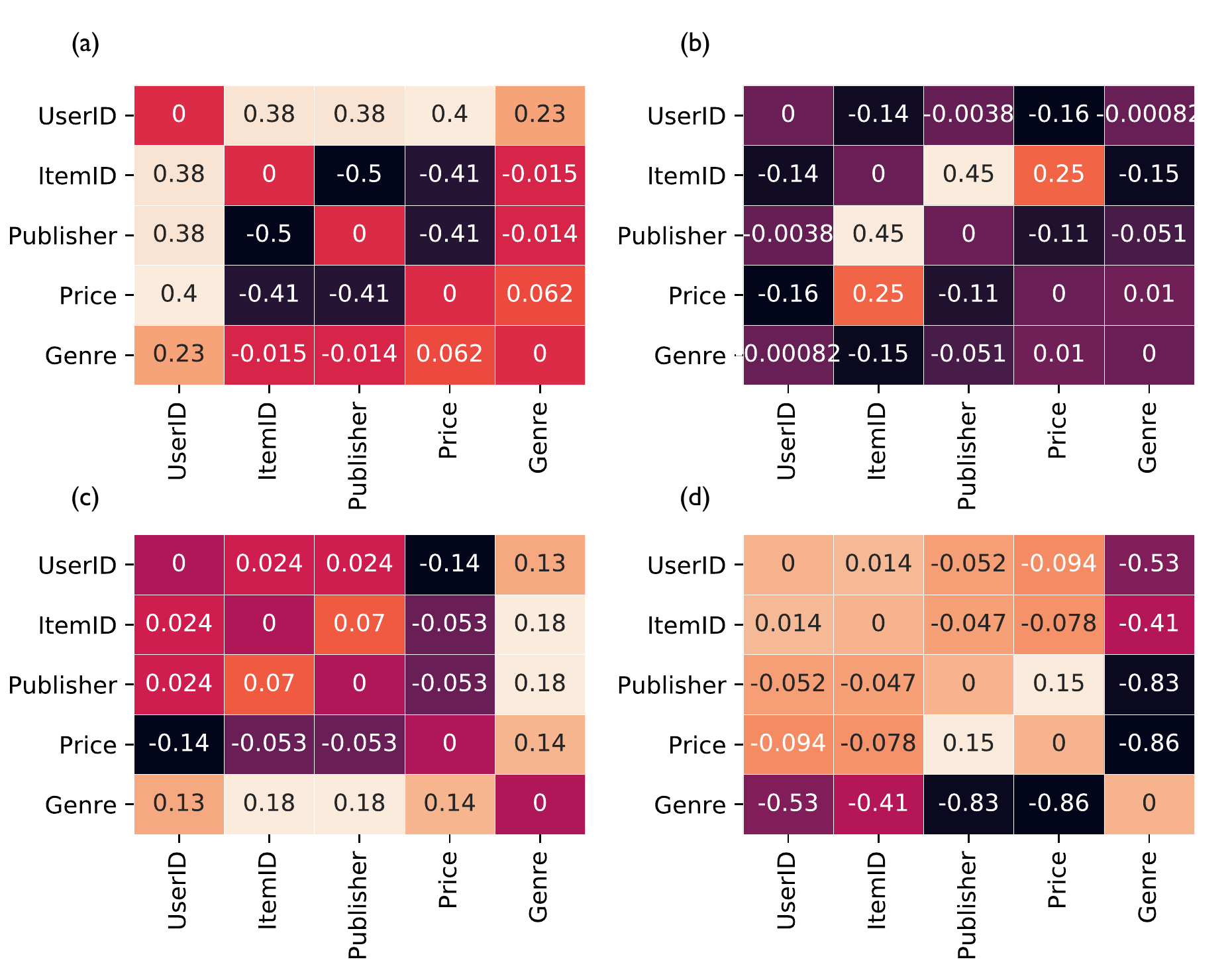}
\end{center}
\caption{Visualization of the heatmap of score functions from LorentzFM for (a) a positive sample and (b) a negative sample, and from FM for the same (c) positive sample and (d) negative sample.} \label{figure:case_study}
\end{figure}

The fine-grained feature interactions are the essence for LorentzFM to outperform deep learning models. Therefore, it is expected that the structure of the feature interactions is capable to explain the human behavior. To demonstrate this, we sample a typical user (user ID ``76561198071045315'') whose almost \textit{all} positive ratings are free in price from the Steam dataset. We then plot the heatmap of feature interaction score of a positive item and a negative item from the test set for both LorentzFM and FM. The result is shown in Figure \ref{figure:case_study}, in which we mask the diagonal elements. From the heatmap we observe the following:

\begin{itemize}
    \item For the positive sample, the game is free of charge. The feature interaction score between ``UserID'' and ``Price'' learned by LorentzFM is highlighted with score 0.4, indicating that price free is strongly preferred by the user, see Figure \ref{figure:case_study} (a). However, as shown in Figure \ref{figure:case_study} (c), the information about this fine structure is missing in the heatmap for FM, because the inner product between the embedding of this ``UserID'' and ``Price'' is -0.14.
    \item For the negative sample, the price of the game is \$14.99. In this case, the score learned by LorentzFM between ``UserID' and ``Price'' is the most negative among others, as shown in Figure \ref{figure:case_study} (b), meaning that the negative sample is recognized mostly due to its price. On the other hand, the inner product between ``UserID'' and ``Price'' is not dominating for FM.
\end{itemize}
This case study shows that compared to FM, the interactions between feature pairs learned from LorentzFM are more informatively related to the underlying facts according to the sign and the magnitude of the pairwise score.

\section{Related Work}\label{section: related_work}

Factorization machine (FM) \cite{FM} and field-aware factorization machine (FFM) \cite{FFM} are two of the most well-known models for both recommendation systems and CTR predictions. FM models the sparse feature interactions by learning latent factors, and FFM introduced field aware factors to increase the model capacity and expressiveness. However, FFM consumes huge memory due to its field awareness and henceforth its scalability is restricted. Therefore, higher-order factorization machine \cite{HOFM} is proposed to model feature interactions beyond quadratic level. 

The research community of CTR prediction recently pays more attention to the deep learning models, due to their strong capabilities in feature extraction. Wide \& Deep model was initially introduced for jointly training wide linear models and deep neural networks to capture both shallow and deep representations. However, these advances still require manual efforts for feature engineering in advance of model training stage. DeepFM \cite{DeepFM}, Neural FM \cite{NFM} and Deep \& Cross Network \cite{DCN} are representative models to combine low-order and high-order feature interactions to improve the performance in CTR in an automatic manner. Product-based Neural Networks \cite{pnn} tries to capture high-order feature interactions by involving a product layer on top of the embedding layer. Apart from the advances in model architectures, attention mechanism, which first appears in neural machine translation models to learn a weighted sum of model outputs, also finds its impact for CTR and recommendation systems. Works that incorporate attention mechanism include Deep Interest Network \cite{DIN}, Attentional Factorization Machine \cite{afm} and AutoInt \cite{AutoINT}. 

Traditional research on recommendation system tries to learn latent factors for user and items based on their preferences \cite{prob_mf,dl_cf}. In these works, learning latent factors by factorizing the interaction matrix is equivalent to using Euclidean inner product of embeddings of user, item and their side information. Our work is inspired by the seminal paper collaborative metric learning \cite{cml} which argues that inner product formulation will result in violation of triangle inequality in Euclidean space. The authors propose to learn not only user-item, but also user-user and item-item distances. This idea is later on generalized to use hyperbolic distance \cite{hyper_rec,scalable_hyper_rec} for recommendation systems, accompanied by the recent advances in hyperbolic representation learning and hyperbolic neural networks \cite{Poincare_embedding,lorentz_model,hyper_entail_cone,Tradeoff_hyperbolic,Lorentz_distance_learning,Hyperbolic_attention_network,radam}. While hyperbolic representation seems eminent over Euclidean representations in many applications, especially for problems with underlying hierarchical structures, previous study focus on learning the distances rather than the violation of triangle inequalities in terms of Lorentz distance. Besides, another type of approach to model feature interactions is to replace the inner product in the FM by circular correlation or circular convolution, as shown in \citet{holo_fm}.

\section{Conclusion}
In this paper, we present a lightweight yet effective model named LorentzFM for recommendation and CTR prediction tasks. We propose a new score function by characterizing if the triangle inequality for Lorentz distance is violated or not in the hyperboloid model. Our model is advantageous with its capability in modeling feature interactions with the underlying geometry, without using any deep learning layers. Empirical experiment results on four real-world benchmark dataset show that LorentzFM achieves the state-of-the-art performance, outperforming multiple existing deep learning baselines, while using fewer training parameters and less training time.

\section{Acknowledgements}
The authors thanks Vivian Tian, Alan Lu, Hua
Yang and Xiaoyuan Wu for their supports, and anonymous reviewers for their valuable comments.

\bigskip
\bibliography{lorentz}
\bibliographystyle{aaai}

\end{document}